\documentclass[12pt]{iopart}

\usepackage{graphicx}
\usepackage{subfigure}

\usepackage{lineno}
\linenumbers

\begin{document}

\title{Optimum performance investigation of LYSO crystal pixels:
A comparison between GATE simulation and experimental data}

\author {Ze Chen}
\address {Institute of Modern Physics, Chinese Academy of Sciences, Lanzhou 730000, China }
\address {University of Chinese Academy of Sciences, Beijing 100049, China}
\ead{chenze@impcas.ac.cn}

\author{Zheng Guo Hu}
\address{Institute of Modern Physics, Chinese Academy of Sciences, Lanzhou 730000, China }
\ead{huzg@impcas.ac.cn}

\author{Jin Da Chen}
\address{Institute of Modern Physics, Chinese Academy of Sciences, Lanzhou 730000, China }
\ead{chenjinda@impcas.ac.cn}

\author{Xiu Ling Zhang}
\address{Institute of Modern Physics, Chinese Academy of Sciences, Lanzhou 730000, China }
\ead{xiuling@impcas.ac.cn}

\author{Zhi Yu Sun}
\address{Institute of Modern Physics, Chinese Academy of Sciences, Lanzhou 730000, China }
\ead{sunzhy@impcas.ac.cn}

\author{Wen Xue Huang}
\address{Institute of Modern Physics, Chinese Academy of Sciences, Lanzhou 730000, China }
\ead{huangwx@impcas.ac.cn}

\author{Jian Song Wang}
\address{Institute of Modern Physics, Chinese Academy of Sciences, Lanzhou 730000, China }
\ead{jswang@impcas.ac.cn}

\author{Zhong Yan Guo}
\address{Institute of Modern Physics, Chinese Academy of Sciences, Lanzhou 730000, China }
\ead{guozhy@impcas.ac.cn}

\author{Guo Qing Xiao}
\address{Institute of Modern Physics, Chinese Academy of Sciences, Lanzhou 730000, China }
\ead{xiaogq@impcas.ac.cn}

\begin{abstract}
Monte Carlo simulation plays an important role in the study of time of flight (TOF)  
positron emission tomography (PET) prototype. As it can incorporate accurate physical modeling of 
scintillation detection process, from scintillation light generation, 
the transport of scintillation photos through the crystal(s), to the conversion of these photons into electronic signals.
The Geant4 based simulation software GATE can provide a user-friendly simulation platform containing the properties needed.
In this work, we developed a dedicated module in GATE simulation tool.
Using this module, we simulated the light yield, energy resolution, 
time resolution of LYSO pixels with the same 
cross-section ($4\times4~mm^{2}$) of different lengths: 5 mm, 10 mm, 15 mm, 20 mm, 25 mm, coupled to a PMT. 
The experiments were performed to validate the GATE simulation results. The results indicate that the best time resoution (484.0$\pm$67.5 ps) and
energy resolution (13.3$\pm$0.4 \% ) could be produced by using pixel with length of 5 mm.
The module can also be applied to other cases for precisely simulating optical photons propagating in scintillators.
\end{abstract}

\section{Introduction}
In the development of time of flight (TOF) positron emission tomography (PET) detectors, 
understanding and optimizing scintillator light collection, energy resolution and time resolution 
is critical for achieving high performance.
Monte Carlo simulations play an important role in guiding research in detector designs, as it can vary many design parameters
much more easily.
The Geant4~\cite{Agostinelli2003250} based simulation platform GATE~\cite{0031-9155-55-6-009}, which provides a user-friendly, scripted interface,
has come into widespread use in the field of nulcear medicine for simulating PET devices.

The widely used LYSO:Ce crystal has high density (7.40 g$\cdot$$cm^{-3}$), high light output (⩾26,000 photons/MeV), 
good energy resolution (∼15\%) and short decay time (40 ns)~\cite{Korzhik2007122},~\cite{0031-9155-47-8-201},
which makes it a good candidate in the study of TOF-PET prototype.
We adapted crystal array which consisted of the LYSO crystal pixels to improve the spatial resolution of our detector modules. 
From the very biginning, we would like to know the performance of LYSO pixels.

The objective of the work is to validate the reliablity of the dedicated module used in GATE simulation tool 
and compare the predicated light yield,
energy resolution and time resolution with experimental results.

\section{Simulation Model and Method}
\begin{figure}
    \centering
    \includegraphics[width=5.0in]{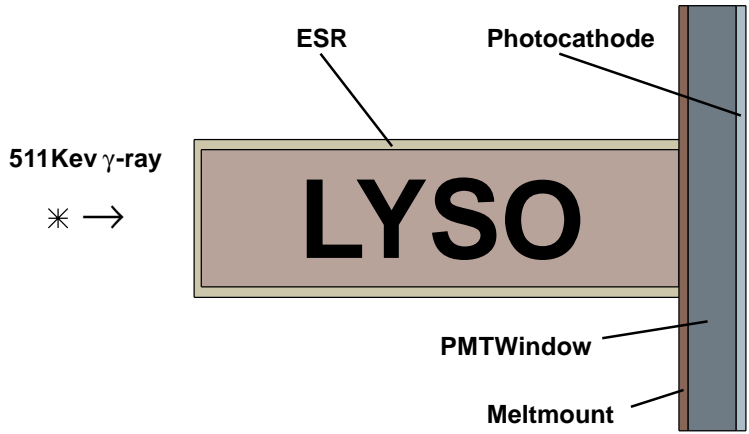}
    \caption{Simplified scheme of the optical volumes and materials used in simulation}
    \label{fig_simulation_scheme}
\end{figure}

The detector is contained within an world volume, which is consist of LYSO crystal coupled to photomultiplier tube (PMT), which is shown in Fig.\ref{fig_simulation_scheme}.
Some elementary properties of the different materials used in the simulation are given in table~\ref{tabl3}. 
Three-part physics process are involved in the simulation: $\gamma$-ray interaction in detector;
the scintillation light emission and light transport inside the detector and the photoelectron generation;
anode signal generation in the PMT and electron process which involes the time discriminator.

\subsection{$\gamma$-ray Generation and Interaction in Detector}
In GATE simulation, 511 KeV $\gamma$-rays imping laterally on different length LYSO crystal pixels at the smallest cross-section
($4\times4 mm^{2}$). In the process of simulation, we change nothing but the length of the LYSO pixel from 5 mm to 25 mm by a step of 5 mm.

\subsection{Scintillation light emission and transport}
The scintillation light emission in GATE was decribed by the total light yield, rise time, decay time, resolution scale,
and bulk light attenuation length,which are listed in table~\ref{tabl4}.

The light transport process in the detector is very complicated. We use the UNIFIED model~\cite{unified1996} to model the refection of the
photons at surfaces between two dielectric materials. In this study, we use ground and ground-back-painted. A ground surface
is assumed to be consist of small micro-facets, whose normals have small angles relative to the average surface normal. The
distribution of these angle is assumed to be Gaussian with mean 0 and standard deviation $\sigma_{\alpha}$. 
The type and surface finish of each of the optical interfaces defined in the simulations have been given in table~\ref{tabl5}.

\subsection{PMT response}

The PMT response of single photoelectron is a current signal that can be described as
a Gaussian pulse~\cite{Bellamy1994}:

\begin{equation}
i_{pe}(t)=\frac{G}{\sqrt{2\pi}\sigma} \exp(-\frac{t^{2}}{2\sigma^{2}}) \label{eq1}
\end{equation}

G = Parameter related to the gain of PMT,

$\sigma$ = Time constant that determines the width of the pulse.

The anode output circuit can be considered as a parallel of a load resistor (R$_{L}$) and
parasitic capacitor (C). The impulse response of the circuit is:

\begin{equation}
h(t)=\frac{1}{C}\exp(-\frac{t}{\tau}) \label{eq2}
\end{equation}

$\tau$=R$_{L}$C

The single photoelectron anode response (v$_{pe}$) was the convolution of the anode current and
the output circuit~\cite{Liu2009}:

\begin{equation}
v_{pe}(t)=\frac{G}{\sqrt{2\pi}\sigma C} \cdot
\int^t_0
\exp(-\frac{t-x}{\tau})\cdot
\exp(-\frac{t^{2}}{2\sigma^{2}}){\rm d}x \label{eq3}
\end{equation}

The PMT output single was the sum of the anode pulses of all the photoelectrons:
\begin{equation}
V(t)=\sum^{N_{pe}}_{i=1}v_{pe,i}(t-t_{pe,i}) \label{eq4}
\end{equation}

\begin{figure}
    \centering
    \includegraphics{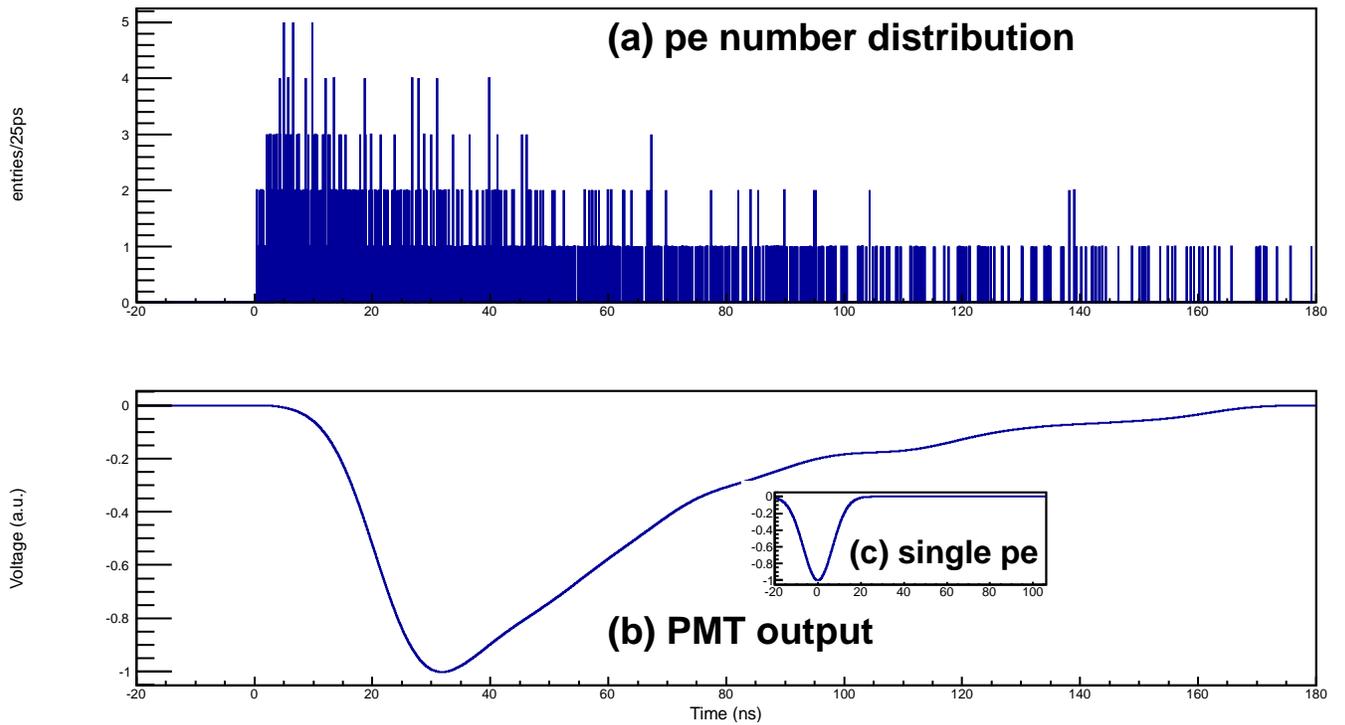}
    \caption{(a) Photoelectrons time distribution on the photocathode, (b) Normalized PMT output signal
     of a $\gamma$-ray event and (c) Normalized PMT response of single photoelectron. Pe = photoelectron.}
    \label{fig_pmt_response}
\end{figure}

Fig.~\ref{fig_pmt_response} shows a simulated LYSO pixel signal of a 511 keV $\gamma$-ray event. 
The data includes the photoelectrons time distribution on the photocathode, the PMT output signal of a gamma event
and the normalized PMT response of single photoelectron(pe), 
Here normalized means that the photoelectron distribution is divided by the height of the spectrum.

\section{Experimental studies}
\subsection{Set up}
\begin{figure}
    \centering
    \includegraphics[width=3.0in]{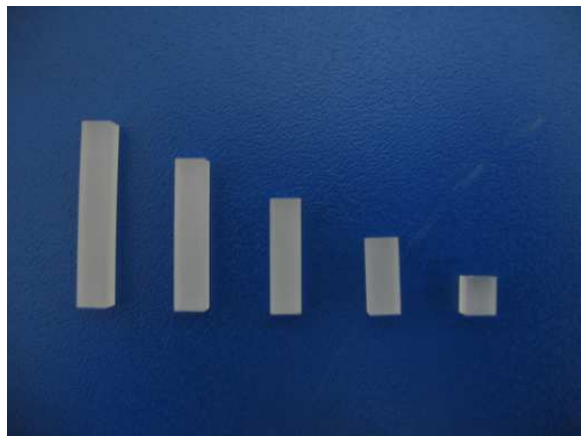}
    \caption{LYSO pixels used in the experiment}
    \label{fig_lyso_pixels}
\end{figure}

The LYSO pixels with various lengths and same cross-section($4\times4 mm^{2}$) are used in the experiment, which are illustated in Fig.~\ref{fig_lyso_pixels}. 
Since the naturally radioaction of Lu(176) in LYSO, a coincidence detection circuit 
was adapted in order to eliminate the LYSO background. The collimator is 8 cm thick lead block with 4 mm diameter holes, which is set on a position system able to move along both the  x- and y-axis. A plane ${}_{}^{22}$Na source was placed in the center of the lead block. The coincidence detector module consists of a $\phi$3 cm LaBr$_3$ crystal, which is produced by Saint-Gobain, coupled to a Hamamatsu R4998 PMT, biased by a high-voltage power supply ORTEC Model 556 with a negative voltage of 1400V. The imaging detector module consists of a LYSO pixel which is producted by Sinoceramics, Inc., coupled to a Photonics XP20D0 PMT, biased by the same high-voltage power supply with a negative voltage of 1000 V. The LaBr$_3$ detector and the imaging detector were placed on two position systems respectively, which could move along both the x- and z-axis. By moving the LaBr$_3$ detector, imaging detector and lead collimator, the collimated 511keV flux could reach both faces of two detector modules.
To increase the collection efficiency for light photons produced by absorption of $\gamma$-rays, the samples under investigation were coated with Enhanced Spectular Reflector(ESR). Optical constact between the scintillation crystal and PMT photocathode is maintained by a optical grease.

\subsection{Coincidence acquisition system}
\begin{figure}
    \centering
    \includegraphics[width=3.0in]{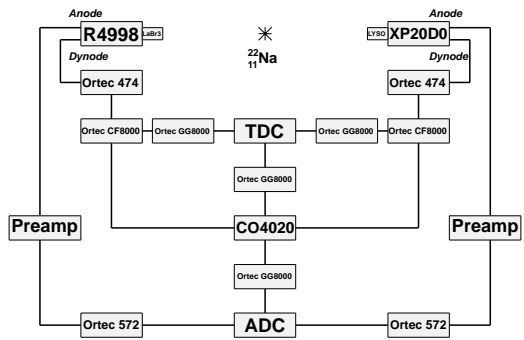}
    \caption{Electronic Configure}
    \label{fig_electronicConfigure}
\end{figure}

The coincidence acquisition system and signal flow is illustated in Fig.~\ref{fig_electronicConfigure}. 
Both the imaging and coincidence detector modules processing units are identical. The dynode signals were transmitted to the coincidence module(Ortec CO4020) via a time filter amplifier(TFA, Ortec 474), a constant fraction discriminator(CFD, Ortec CF8000) and a delay(Ortec GG8000). The discriminated signals from CFD were also fed into TDC module(Philips 7187). The anode signal flows were sent to a amplifier(Ortec 572) via a preamplifier. The amplified signals were fed into ADC module(Philips 7164).

\section{Comparison between simulation and experiment}

\subsection{Energy spectrum}
Fig.~\ref{fig_energy_spectrum} shows the energy spectra of LYSO pixels with different length. All spectrums are normalized
such that the full-energy peaks have equal heights. Both compton platform and photo peak are well described by the GATE simulation with different pixel lengths.
The low energy part of measured spectrum was truncated due to the CFD setting.
Slight discrepancies between the experimental and simulated energy photopeak is observed but these difference are not significant.

\subsection{Light yield}
\begin{figure}
    \centering
    \includegraphics[width=4.0in]{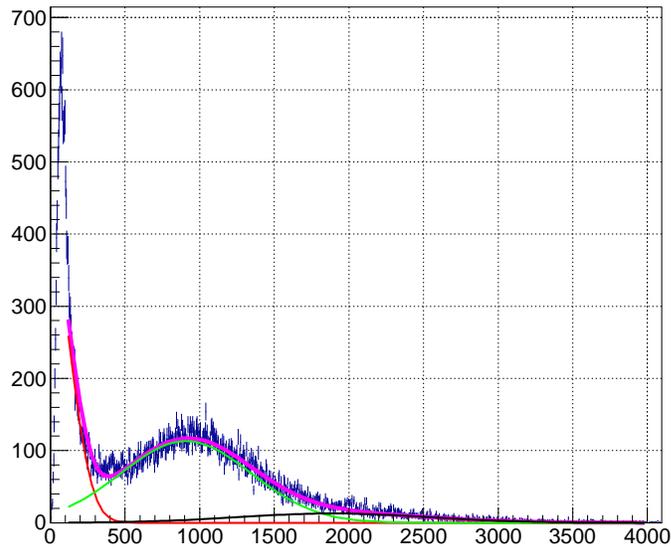}
    \caption{Single Photoelectron Spectrum: magenta line represents the total fit to the spectrum, red line presents gaussian fit to the pedestal, green line represents gaussian fit to the single photo-electron peak and black line represents gaussian fit to the two photo-electrons peak.}
    \label{fig_spe}
\end{figure}

For quantitative analysis, the simulated light output was derived by integrating the total area under the histogram of photoelectrons
from time zero to a time point which was three times the decay constant of scintillation decay (Fig.~\ref{fig_pmt_response}(a)).
The measured absolute light yield of the scintillation pixel under investigatiion is determined by comparing the postion of the full energy peak in the $^{22}$Na 
spectra (511 keV) to the position of center-of-gravity of the single-photoelectron spectrum~\cite{Bertolaccini1968}, which is shown in Fig.~\ref{fig_spe}.
We use multi-gaussian fit to the spectrum to find mean value of the single photo-electron peak, which can be regarded as centr-of-gravity.

\begin{figure}
    \centering
    \includegraphics{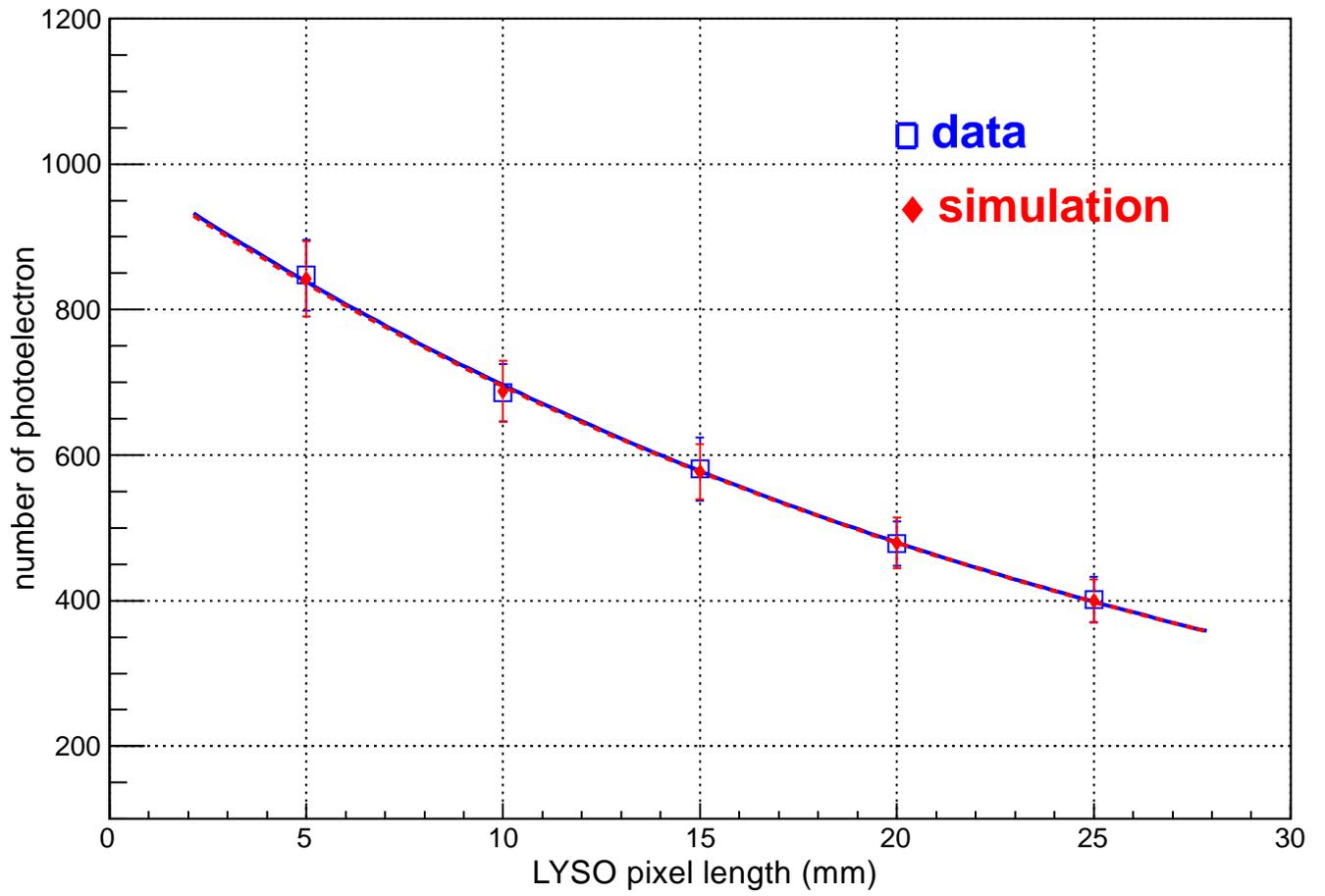}
    \caption{Experiment (box) and simulation (diamond) light yield. Error bars represent one standand deviation of uncertainty.}
    \label{fig_light_yield}
\end{figure}

Fig.~\ref{fig_light_yield} shows absolute light yield as a function of the crystal length. The error bars represent one standand deviation of uncertainty. A exponential fit was used to abtain the attenuation length. The simulated effective attenuation length is 26.72$\pm$0.06 mm, less than the bulk length used above. The experimental attenuation is found to be equal to 27.05$\pm$0.06 mm. 
Good agreement is found between the simulation and experiment.

\subsection{Energy resolution}
\begin{figure}
    \centering
    \includegraphics{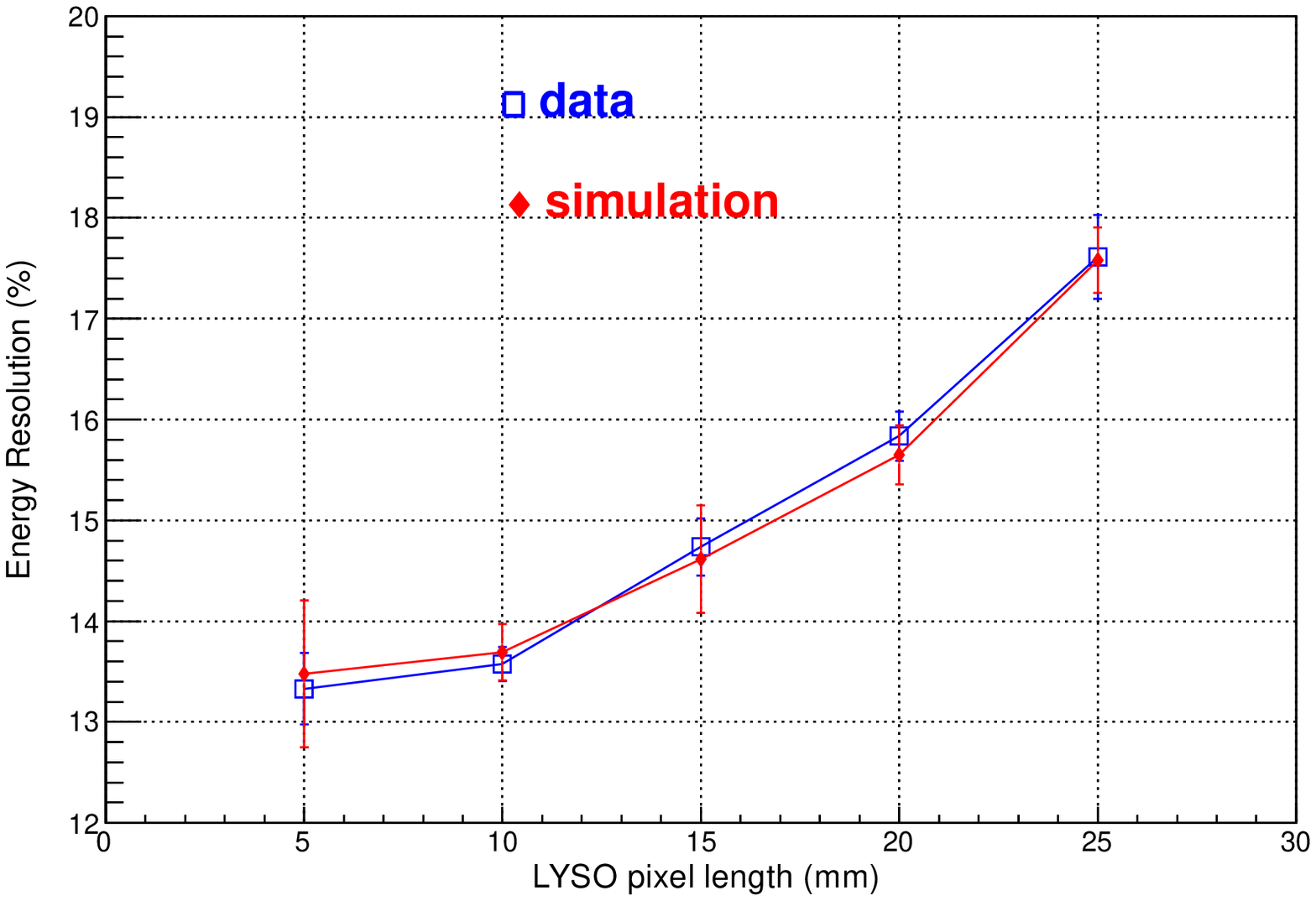}
    \caption{Experiment (box) and simulation (diamond) energy resolution. Error bars represent the statistical errors.}
    \label{fig_energy_resolution}
\end{figure}

For each energy spectrum, the energy resolution was obtained from a gaussian fit to the photopeak. 
Fig.~\ref{fig_energy_resolution} shows the measured and simulated energy resolution of LYSO pixels with different lengths.  
The error bars did not comprise systematics errors, but only the statistical errors,
which can be calculated from statistical uncertainties of the fitted mean and sigma using propagation of errors formula.
The larger error bar in pixel of length 05 mm reflects the fact, shorter pixel has low-efficiency to detect $\gamma$-ray.

Modeled and measured energy resolutiion values agreed very well with discrepancies
 limited to the range of -1.18 \% to +1.12 \% and an average absolute difference of 0.83 \%.
This excellent agreement indicates that the energy resolutions observed in pixels with different lengths can be reproduced 
and attributed entirely to statistical uncertaintions in the number of photoelectrons detected  at the photocathode.

\begin{figure}
    \centering
    \subfigure[]
    {
        \includegraphics[width=2.0in]{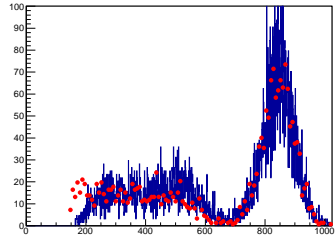}
        \label{fig_energy_spectrum_sub_1}
    }
   \\
    \subfigure[]
    {
        \includegraphics[width=2.0in]{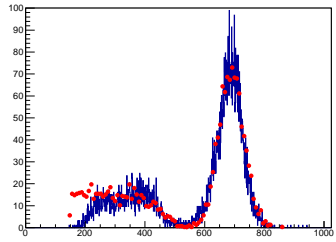}
        \label{fig_energy_spectrum_sub_2}
    }
    \subfigure[]
    {
        \includegraphics[width=2.0in]{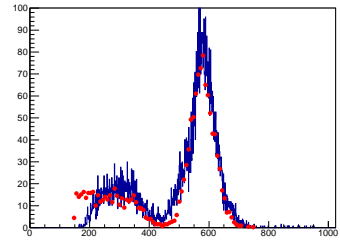}
        \label{fig_energy_spectrum_sub_3}
    }
        \subfigure[]
    {
        \includegraphics[width=2.0in]{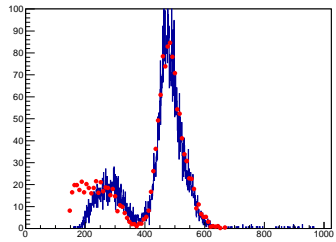}
        \label{fig_energy_spectrum_sub_4}
    } 
        \subfigure[]
    {
        \includegraphics[width=2.0in]{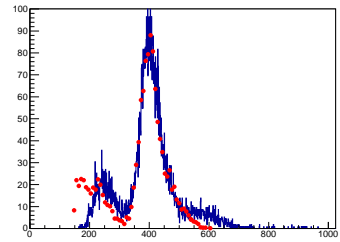}
        \label{fig_energy_spectrum_sub_5}
    }
    \caption{Experiment (blue line) and simulation (red dot) Energy Spectrum of different LYSO pixel lengths: (a) L=05 mm,(b) L=10 mm,(c) L=15 mm,(d) L=20 mm,(e) L=25 mm.}
    \label{fig_energy_spectrum}
\end{figure}

The tail pile-up is observed in Fig.~\ref{fig_energy_spectrum}(c). This may at least partially be caused by the undershoots from preceding pulse during acquication.
In Fig.~\ref{fig_energy_spectrum}(e), a long tail in high energy part of experiment spectrum is obvious, when compared to simulation one.
We find the similar tail in simulation by using a shorter bulk length (i.e. 30 mm) in crystal setting. 
From this point of view, the effective attenuation length in pixel of length 25 mm should be less than 26.5 mm, which may be caused by non-uniformity of lyso pixel.

\subsection{Time resolution}
For both simulation and experiment, we use constant fraction discrimination (CFD) as the time pick-off method. 
In the CFD implemention of the module, 
the original pulse histogram is attenuated by a factor and then added to an inverted version of the raw signal histogram with a delay. 
The time information can be derived by finding the position of zero-crossing using interpolation.
We can extract the time resolution $\Delta$t$_{s}$ from a gaussian fit to time information histogram.
Since PMT transit-time-dispersion (TTS) is independent of the scintillation and photo-electron conversion,
so the detector time resolution $\Delta$t$_{det}$ is calculated with TTS (200 ps \cite{Stefan2012}) and $\Delta$t$_{s}$ as~\cite{0031-9155-52-4-016}

\begin{equation}
\Delta t_{det}=\sqrt{(\Delta t_{s})^{2} + (TTS)^{2}}  \label{eq5}.
\end{equation}

\begin{figure}
    \centering
    \includegraphics{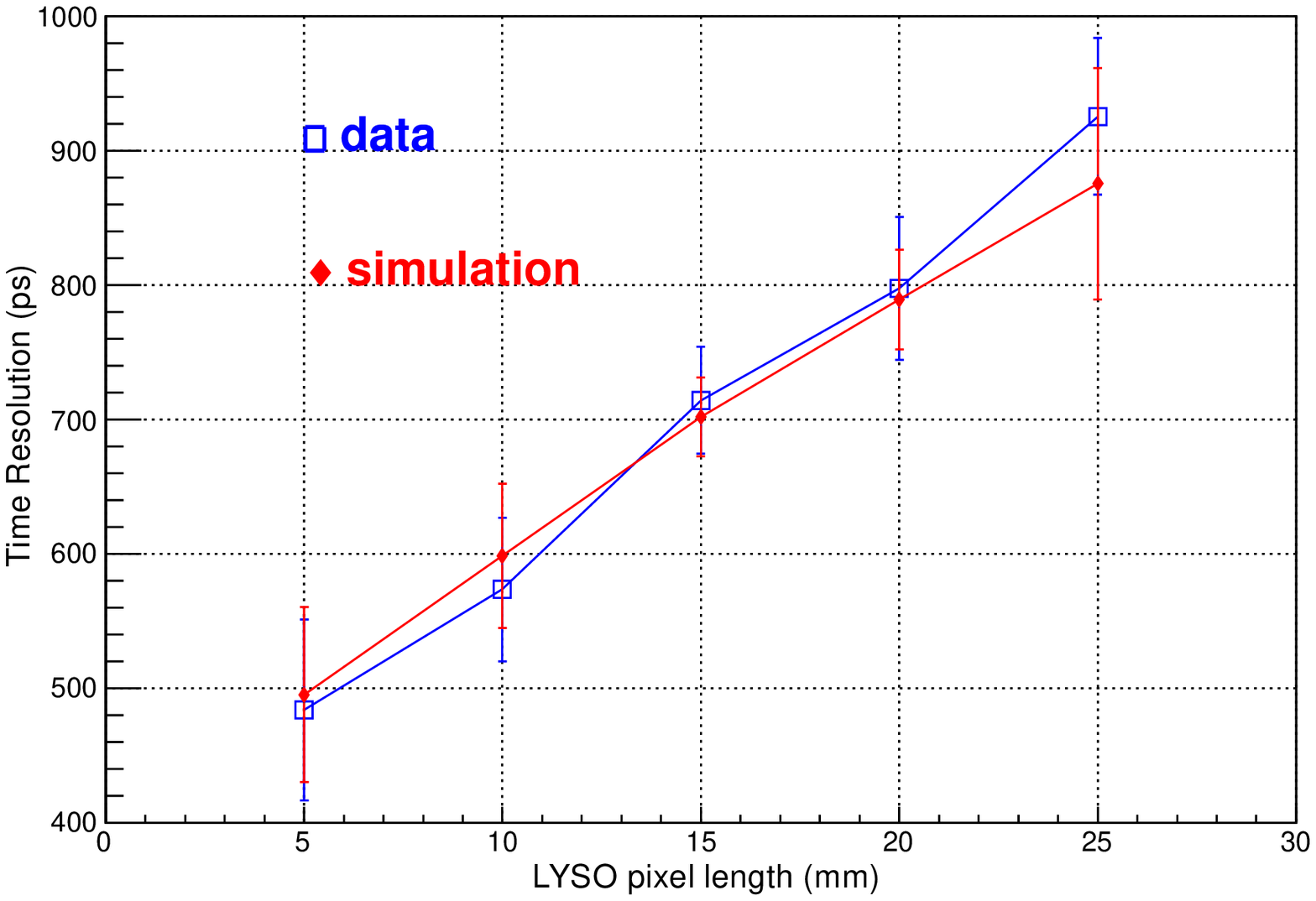}
    \caption{Experiment (box) and simulation (diamond) time resolution. Error bars represent the statistical errors.}
    \label{fig_time_resolution}
\end{figure}

Fig.~\ref{fig_time_resolution} shows the time resolution abtained by simulation and experiment. 
The error bars only comprise the statistical errors. Good agreement is found within the simulation and experiment errors.

\section{Conclusions and future work}
We have developed a specified module in GATE simulation software,
which has proved to be reliable in fully simulating the optical photon processes in the pixel geometry.
This module can also easily be used for block geometry, which is consist with matrix of pixels.

For future work, we will incorporate some mathematical models to accelerate simulation process and simulate the
optical processes in the Anger-logic based detector to generate the spatial distribution information,
from which light sharing can be investigated.

\section*{Acknowledgements}
This work is supported by National Natural Science Foundation of China (11205222), West Light Foundation of the Chinese Academy of Sciences (210340XBO) ，National Major Scientific Instruments and Equipment Development Projects (2011YQ12009604)，Youth Innovation Promotion Association, CAS（201330YQO）

\section*{References}
\bibliography{myrefs}

\begin{table}[htbp]
\caption{\label{tabl3} Properties of the materials used in the simulation.}
\begin{indented}
\item[]\begin{tabular}{@{}llll}
\br
&Chemical&Density&Refractive\\
Material&composition&(g\,cm$^{-3}$)&index\\
\mr
Air&N$_{0.76}$O$_{0.23}$&$1.29 \times 10^{-3}$&1.00 \\
LYSO&Lu$_{2}$Si$_{1}$O$_{5}$&7.40&1.82\\
Meltmount&C$_{1}$H$_{1}$O$_{1}$&1.00&1.58 \\
PMTWindow&C$_{1}$H$_{1}$O$_{1}$&1.00&1.52 \\
Photocathode&Aluminium&2.70&- \\
\br
\end{tabular}
\end{indented}
\end{table}

\begin{table}[htbp]
\caption{\label{tabl4} Optical Properties of LYSO.}
\begin{indented}
\item[]\begin{tabular}{@{}lllll}
\br
Light yield&Rise time&Decay time&Resolution scale&Bulk light attenuation length\\
\mr
26000 photos/Mev&0.09 ns\cite{Stefan2012}&40 ns&6.8&40 mm\\
\br
\end{tabular}
\end{indented}
\end{table}

\begin{table}[htbp]
\caption{\label{tabl5} Type and surface finish of each of the optical interfaces defined in the simulations}
\begin{indented}
\item[]\begin{tabular}{@{}lllll}
\br
Name&type&finish&$\sigma_{\alpha}$\\
\mr
LYSO-Frontside&dielectic\_dielectric&groundbackpainted&0.1 degrees\\
LYSO-Foursides&dielectic\_dielectric&groundbackpainted&4.0 degrees\\
LYSO-Meltmount&dielectic\_dielectric&ground&0.1 degrees\\
Meltmount-PMTWindow&dielectic\_dielectric&ground&0.1 degrees\\
PMTWindow-Photocathode&dielectic\_metal&ground&0.0  degrees\\
\br
\end{tabular}
\end{indented}
\end{table}

\end{document}